\documentclass[conference,10pt]{IEEEtran}
\usepackage{amsmath,amssymb}
\usepackage{tikz,pgfplots}
\usepackage{algorithm}
\usepackage{algorithmic}
\usepackage{subcaption}
\usepackage{cite}
\usepackage{url}

\title{ParamNet: A Multi-Layer Parametric Network for Joint Channel Estimation and Symbol Detection}

\makeatletter
\newcommand{\linebreakand}{%
  \end{@IEEEauthorhalign}
  \hfill\mbox{}\par
  \mbox{}\hfill\begin{@IEEEauthorhalign}
}
\makeatother

\author{\IEEEauthorblockN{Vincent Choqueuse}
\IEEEauthorblockA{\textit{Lab-STICC, UMR CNRS 6285}\\
\textit{ENIB, Technopole Brest-Iroise} \\
29238 Brest Cedex 3, France \\
choqueuse@enib.fr}
\and
\IEEEauthorblockN{Alexandru Frunza}
\IEEEauthorblockA{\textit{Lab-STICC, UMR CNRS 6285}\\
\textit{ENIB, Technopole Brest-Iroise} \\
29238 Brest Cedex 3, France \\
frunza@enib.fr}
\and
\IEEEauthorblockN{Adel Belouchrani}
\IEEEauthorblockA{\textit{Electrical Engineering Department} \\
\textit{Ecole Nationale Polytechnique}\\
Algiers, Algeria \\
adel.belouchrani@g.enp.edu.dz}
\linebreakand %
\IEEEauthorblockN{Stéphane Azou}
\IEEEauthorblockA{\textit{Lab-STICC, UMR CNRS 6285}\\
\textit{ENIB, Technopole Brest-Iroise} \\
29238 Brest Cedex 3, France \\
azou@enib.fr}
\and
\IEEEauthorblockN{Pascal Morel}
\IEEEauthorblockA{\textit{Lab-STICC, UMR CNRS 6285}\\
\textit{ENIB, Technopole Brest-Iroise} \\
29238 Brest Cedex 3, France \\
morel@enib.fr}
}

\pgfplotsset{compat=1.17}
\begin{document}
\maketitle
\begin{abstract}
This paper proposes a parametric-based network architecture for joint channel estimation and data detection in communications systems with hardware impairments. This architecture is composed of a data-augmented layer, a custom soft thresholding function, and several linear layers modeling the effect of channel effects and hardware impairments. In the proposed network, the soft thresholding function softly constrains the detected data to be within the considered constellation. The latter depends only on one one parameter that is optimized during training. The benefit of the proposed approach is illustrated through a communication chain corrupted by multiple impairments and noises.
\end{abstract}

\section{Introduction}

In communications systems, the use of advanced coding schemes and modulation formats makes the performance of the system quite sensible to signal distortion occurring at the physical layer. These distortions include the effect of the propagation channel, non-ideal synchronization, IQ imbalance, and carrier impairments. In the literature, a large number of digital algorithms have been developed for hardware impairment compensation and data detection~\cite{VAN97,TUG00,VAL01,TAR05,TUB05,ANT08,SPA11,MEH12,SAL14,LES17,FRU21}. These algorithms are mainly based on parametric signal processing approaches. Recently, several studies have investigated the use of Deep Learning (DL) techniques to address some physical-layer problems \cite{ZAP19,O2017}. In particular, many studies have focused on the use of model-based DL techniques for Multiple-Input Multiple-Output (MIMO) detection~\cite{TAK19,SAM19,KHA20,SHL20}. The main drawback of these approaches is that they usually require perfect channel state information. To address this issue, a DL network addressing the joint channel estimation and data detection problem has been proposed in~\cite{HE20}. As this technique does not assume any apriori model for the channel matrix, it often requires a large training database and is not suited for time-varying channels.

In this paper, we focus on the joint channel estimation and data detection problem in frequency-selective Single-Input Single-Output (SISO) communications. By modeling the channel as a cascade of parametric models, we propose a new parametric multi-layer network for channel estimation and data detection which is well suited for time-varying channels. This network can be trained using a small number of pilot symbols. Following the principle of Iterative thresholding algorithms, we propose to include a non-linear soft thresholding layer in the structure of the parametric network. This layer is based on the optimal denoiser for Gaussian noise that has been introduced in \cite{KHA20}. 

The paper is organized as follows. Section \ref{sec:1} describes the problem statement, Section \ref{sec:2} presents the proposed network architecture, and Section \ref{sec:3} illustrates the benefit of our approach together with the obtained simulation results.

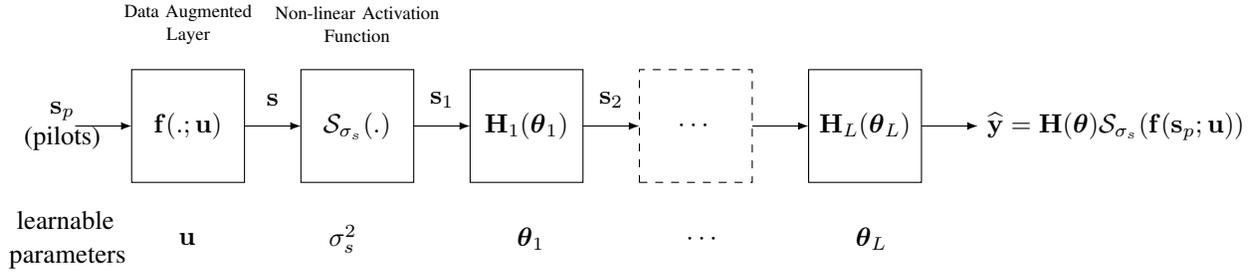
\begin{figure*}[!t]
\centering
\begin{tikzpicture}
\node[right, text centered,text width=4em, ] at (-4.25,1) {$\mathbf{s}_p$\\(pilots)};
\draw[->,>=latex] (-3.25,1)--(-2.5,1);
\draw (-2.5,0.25) rectangle (-1,1.75) node[pos=.5,text width=4em, text centered](l1) {$\mathbf{f}(.;\mathbf{u})$} ;
\draw[->,>=latex] (-1,1)--(-0.25,1) node[midway,yshift=1em]{$\mathbf{s}$} ;
\draw (-0.25,0.25) rectangle (1.25,1.75) node[pos=.5,text width=4em, text centered](l1) {$\mathcal{S}_{\sigma_s}(.)$} ;
\draw[->,>=latex] (1.25,1)--(2,1) node[midway,yshift=1em]{$\mathbf{s}_1$} ;
\draw (2,0.25) rectangle (3.5,1.75) node[pos=.5,text width=4em, text centered](l1) {$\mathbf{H}_1(\boldsymbol \theta_1)$} ;
\draw[->,>=latex] (3.5,1)--(4.25,1) node[midway,yshift=1em]{$\mathbf{s}_2$} ;
\draw[dashed] (4.25,0.25) rectangle (5.75,1.75) node[pos=.5,text width=4em, text centered] {$\cdots$};
\draw[->,>=latex] (5.75,1)--(6.5,1);
\draw (6.5,0.25) rectangle (8,1.75) node[pos=.5,text width=4em, text centered]  (l3){$\mathbf{H}_L(\boldsymbol \theta_L)$};
\draw[->,>=latex] (8,1)--(8.75,1) node[right] {$\widehat{\mathbf{y}}=\mathbf{H}(\boldsymbol \theta)\mathcal{S}_{\sigma_s}(\mathbf{f}(\mathbf{s}_p;\mathbf{u}))$};
\node at (-1.75,2.5) {\scriptsize Data Augmented};
\node at (-1.75,2.2) {\scriptsize Layer};
\node at (0.5,2.5) {\scriptsize Non-linear Activation};
\node at (0.5,2.2) {\scriptsize Function};
\node[right] at (-4.15,-0.25) {learnable};
\node[right] at (-4.25,-0.75) {parameters};
\node[right] at (-2.,-0.5) {$\mathbf{u}$};
\node[right] at (0,-0.5) {$\sigma_s^2$};
\node[right] at (2.5,-0.5) {$\boldsymbol \theta_1$};
\node[right] at (4.75,-0.5) { $ \cdots$ };
\node[right] at (7,-0.5) {$\boldsymbol \theta_L$};
\end{tikzpicture}

\caption{Proposed ParamNet network architecture. The network is composed of a data augmented layer, $\mathbf{f}(.;\mathbf{u})$, a non-linear activation function, $\mathcal{S}_{\sigma_s}(.)$, and $L$ linear layers $\mathbf{H}_l(\boldsymbol \theta_l)$ modeling the hardware impairments and channel effects.}\label{sys1}
\end{figure*}

\section{Problem Statement}\label{sec:1}

\subsection{Signal Model}
Let us consider a communications channel with $N$ complex-valued transmitted 
symbols $\mathbf{s}_c=[s_{c}[0],\cdots,s_{c}[N-1]]^T$ and $N$ complex-valued received samples $\mathbf{y}_c=[y_{c}[0],\cdots,y_{c}[N-1]]^T$, where $^T$ corresponds to the vector transpose. Let us denote the augmented real-valued transmitted symbols and received samples by $\mathbf{s}=[\Re e(\mathbf{s}_c^T),\Im m(\mathbf{s}_c^T)]^T$ and $\mathbf{y}=[\Re e(\mathbf{y}_c^T),\Im m(\mathbf{y}_c^T)]^T$, respectively. By assuming a linear channel, the augmented received samples can be expressed by the following linear model:
\begin{align}
\mathbf{y} = \mathbf{H}(\boldsymbol \theta)\mathbf{s}+\mathbf{b},\label{eq1}
\end{align}
where $\mathbf{H}(\boldsymbol \theta)$ is a $2N\times 2N$ global transfer matrix, $\boldsymbol \theta$ corresponds to the unknown channel parameters, and $\mathbf{b}\sim \mathcal{N}(\mathbf{0},\sigma^2_b \mathbf{I})$ is an additive noise component with zero mean and variance $\sigma^2_b$. In communications systems, the transmitted data usually belong to a finite alphabet. This property implies that each augmented symbol $s[n]$ belongs to a set $\mathcal{M}=\{M_1,\cdots,M_K\}$ composed of $K$ real-valued symbols. Finally, we consider in this study that the global transfer matrix can be decomposed as a multi-layer channel as follows
\begin{align}
\mathbf{H}(\boldsymbol \theta)=\mathbf{H}_L(\boldsymbol \theta_L)\times \cdots \times \mathbf{H}_1(\boldsymbol \theta_1),
\end{align}
where $\mathbf{H}_l(\boldsymbol \theta_l)$ corresponds to the transfer matrix of the $l^{th}$ layer and depends on the unknown parameters $\boldsymbol \theta_l$. Note that this multi-layer model arises naturally in communications systems corrupted by multiple linear impairments or effects such as transmitter or receiver IQ imbalance, Carrier Frequency Offset, Phase Noise, frequency selective channels, \emph{etc}.

\subsection{Joint Estimation and Detection Problem}

The objective of this paper is to jointly estimate $\boldsymbol \theta$ and $\mathbf{s}\in \mathcal{M}^{2N}$ from $\mathbf{y}$.
For Gaussian noise, this optimization problem can be formulated as $\min_{\boldsymbol \theta,\mathbf{s}\in \mathcal{M}^{2N}}\|\mathbf{y}-\mathbf{H}(\boldsymbol \theta)\mathbf{s}\|^2_2
$. This problem is challenging since $\boldsymbol \theta$ is unknown and $\mathbf{s}$ belongs to a discrete alphabet. To simplify this problem, one solution is to consider that $N_x$ complex-valued pilot symbols, $\mathbf{s}_{c}$, are known at the receiver. By using the knowledge of the real and imaginary parts of the pilot symbols, denoted by the augmented vector $\mathbf{s}_p=[\Re e(\mathbf{s}_{c}^T),\Im m(\mathbf{s}_{c}^T)]^T$, the optimization problem can be reformulated as 
\begin{align}
\min_{\boldsymbol \theta,\mathbf{u}\in \mathcal{M}^{2N_u}}\|\mathbf{y}-\mathbf{H}(\boldsymbol \theta)\mathbf{f}(\mathbf{s}_p;\mathbf{u})\|^2_2,
\end{align}
where $\mathbf{u}$ is an unknown vector composed of $2N_u$ symbols, and $\mathbf{s}=\mathbf{f}(\mathbf{s}_p;\mathbf{u})$ is a data-augmented function that takes $2N_x$ real-valued symbols and returns $2N=2(N_x+N_u)$ real-valued symbols. 

To estimate $\boldsymbol \theta$ and $\mathbf{u}$, two approaches can be employed. In the first approach, the alphabet of $\mathbf{u}$ can be simply neglected by assuming $\mathbf{u}\in \mathbb{R}^{2N_u}$. In the second approach, the alphabet can be enforced using a projected gradient algorithm by mapping periodically $\mathbf{u}$ to the closest point in $\mathcal{M}^{2N_u}$. Even if this second approach has the distinct advantage of exploiting the data constellation, the influence of projection errors in the medium-SNR region cannot be completely excluded. Furthermore, during training, the use of a hard-thresholding function makes it impossible to backpropagate the gradient to the data augmented layer since the function derivative is $0$ almost everywhere. To address these two issues, we propose a new network architecture composed of a soft projection layer.

\section{Proposed Multi-Layer Network}\label{sec:2}

The structure of the proposed Multi-Layer parametric network is presented in Fig.~\ref{sys1}. This section describes the mathematical models of each layer and presents the associated learning algorithm.

\subsection{Layer Models}

In this section, the real-valued inputs and outputs of each layer are denoted $\mathbf{x}_i$ and $\mathbf{x}_o$, respectively.\\

\subsubsection{Data Augmented Layer}

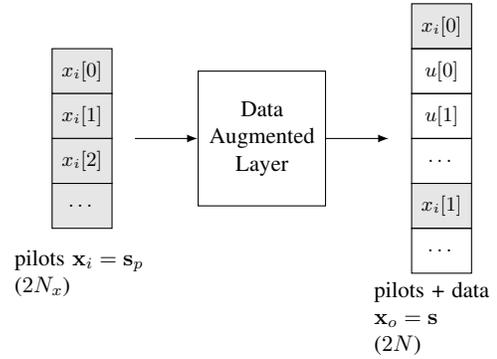
\begin{figure}[!h]
\centering
\begin{tikzpicture}[scale=0.85, every node/.style={scale=0.85}]
 \tikzset{cell/.style={rectangle,
                        text width=2em,
                        minimum height=2em,
                        text centered,
                        draw}}
\node[cell,fill=black!10] (l1) {\small $x_i[0]$} ;
\node[cell,fill=black!10,below of=l1,node distance=2em] (l2) {\small $x_i[1]$} ;
\node[cell,fill=black!10,below of=l2,node distance=2em] (l3) {\small $x_i[2]$} ;
\node[cell,fill=black!10,below of=l3,node distance=2em] (l4) {\small $\cdots$} ;
\node[below of=l4,node distance=3em,text width=6em] (l5) {pilots $\mathbf{x}_i=\mathbf{s}_p$\\($2N_x$)};
\node[rectangle, draw, right of=l2,text width=5em, text centered,yshift=-1em,node distance=8em,minimum height=6em] (r1) {Data Augmented Layer} ;
\node[cell,fill=black!10,right of=l1,node distance=16em,yshift=2em] (x1) {\small $x_i[0]$} ;
\node[cell,below of=x1,node distance=2em] (x2) {\small $u[0]$} ;
\node[cell,below of=x2,node distance=2em] (x3) {\small $u[1]$} ;
\node[cell,below of=x3,node distance=2em] (x4) {\small $\cdots$} ;
\node[cell,fill=black!10,below of=x4,node distance=2em] (x5) {\small $x_i[1]$} ;
\node[cell,below of=x5,node distance=2em] (x6) {\small $\cdots$} ;
\node[below of=x6,node distance=3em,text width=6em] (l5) {pilots + data\\ $\mathbf{x}_o=\mathbf{s}$\\($2N$)};
\node[left of=r1,node distance=6em] (r0){};
\node[right of=r1,node distance=6em] (r2){};
\draw[->,>=latex] (r0)--(r1);
\draw[->,>=latex] (r1)--(r2);
\end{tikzpicture}
\caption{Illustration of the Data Augmented Layer\label{layer1}. The learnable parameters correspond to the data  $\mathbf{u}=[u[0],u[1],\cdots]^T$.}
\end{figure}

The purpose of the data augmented layer is to insert some data $\mathbf{u}$ into the input vector $\mathbf{x}_i$ (see Fig.~\ref{layer1}). Mathematically, the input-output relationship of this layer can be expressed by
\begin{align}
\mathbf{x}_o=\mathbf{f}(\mathbf{x}_i;\mathbf{u})=\mathbf{P}_0\mathbf{x}_i+\mathbf{P}_1\mathbf{u},
\end{align}
where the layer trainable parameters correspond to the unknown real-valued vector $\mathbf{u}$. The matrices $\mathbf{P}_0$ and $\mathbf{P}_1$ correspond to \emph{allocation matrices} of size $2N\times 2N_x$ and $2N\times 2N_u$, respectively. An allocation matrix is a matrix with entries from $\{0,1\}$ that only contains a single one in each row.

\subsubsection{Non-linear Activation Function}
The non-linear activation layer is based on the soft thresholding operator
\begin{align}
\mathbf{x}_o=\mathcal{S}_{\sigma_s}(\mathbf{x}_i).
\end{align}
The purpose of the function $\mathcal{S}_{\sigma_s}(.)$ is to mimic a soft projection into the data constellation. To this end, we propose to use the optimal denoiser for Gaussian noise, which has been recently used in MIMO detection problems~\cite{KHA20}. By assuming  an i.i.d. Gaussian distribution with diagonal covariance matrix $(\sigma_s^2/2) \mathbf{I}$ for the error $\mathbf{x}_i-\mathbf{s}$ and a uniform distribution over $\mathcal{M}$ for the data symbols $\mathbf{s}$, the $n^{th}$ output of the optimal denoiser can be expressed as \begin{align}
[\mathcal{S}_{\sigma_s}(\mathbf{x}_i)]_n = \frac{ \sum_{s\in \mathcal{M}} s e^{-\frac{1}{\sigma_s^2}(x_i[n]-s)^2}}{\sum_{s\in \mathcal{M}} e^{-\frac{1}{\sigma_s^2}(x_i[n]-s)^2}}
\end{align}
This layer contains only one learnable parameter called the \emph{data noise variance} $\sigma_s^2$. 

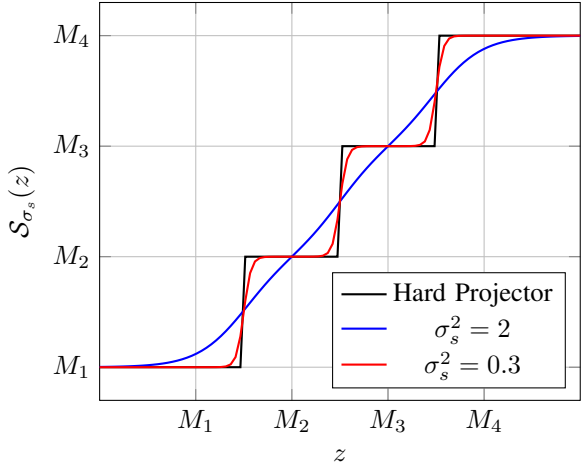
\begin{figure}[!t]
\centering
\begin{tikzpicture}
[declare function = {
    S(\x,\var) = (-3*exp(-((x+3)^2)/\var)-1*exp(-((x+1)^2)/\var)+1*exp(-((x-1)^2)/\var)+3*exp(-((x-3)^2)/\var))/(exp(-((x+3)^2)/\var)+exp(-((x+1)^2)/\var)+exp(-((x-1)^2)/\var)+exp(-((x-3)^2)/\var));
    }
]
\begin{axis}[xlabel=$z$,ylabel=$\mathcal{S}_{\sigma_s}(z)$,xmin=-5,xmax=5,grid=both,width=0.9\linewidth,legend pos=south east,xtick={-3,-1,1,3},xticklabels={$M_1$,$M_2$,$M_3$,$M_4$},
ytick={-3,-1,1,3},yticklabels={$M_1$,$M_2$,$M_3$,$M_4$}]
\addplot[black, thick,samples=100]{S(x,0.01)};
\addplot[blue, thick,samples=100]{ S(x,2)};
\addplot[red, thick,samples=100]{S(x,0.3)};
\legend{Hard Projector, $\sigma_s^2=2$, $\sigma_s^2=0.3$}
\end{axis}
\end{tikzpicture}
\caption{Soft projection into $\mathcal{M}=\{M_1,M_2,M_3,M_4\}$.}\label{fig_soft}
\end{figure}

For illustration purpose, Fig.~\ref{fig_soft} depicts the function $\mathcal{S}_{\sigma_s}(.)$ for different values of $\sigma_s^2$. Note that the data noise variance $\sigma_s^2$ is different from the additive noise variance $\sigma_b^2$ in \eqref{eq1}. The data noise variance $\sigma_s^2$ allows to alternate between the exploration and exploitation of the data vector $\mathbf{u}$ during training. For example, the exploration is encouraged when $\sigma_s^2= 2$ since  $\mathcal{S}_{\sigma_s}(.)$ is nearly linear between $M_1$ and $M_4$, while the exploitation is encouraged when $\sigma_s^2=0.3$ since this function is close to a threshold detector.

\subsubsection{Parametric SISO Channel Layers}

This section presents a list of linear layers commonly encountered in communications systems. While this list is not exhaustive and only focuses on SISO layers, it can be easily adapted to more specific problems or MIMO systems. For SISO layers, the augmented outputs can be expressed as $\mathbf{x}_{o}=\mathbf{H}_l(\boldsymbol \theta_l)\mathbf{x}_{i}$, where the augmented inputs $\mathbf{x}_{i}=[\Re e(\mathbf{x}_{c,i}^T),\Im m(\mathbf{x}_{c,i}^T)]^T$ contains the real and imaginary parts of the complex-valued input vector $\mathbf{x}_{c,i}=[x_{c,i}[0],\cdots,x_{c,i}[N-1]]^T$. Herein, the mathematical models are expressed in scalar form for sake of simplicity.

\begin{itemize}
\item IQ Imbalance Layer: The output of an IQ imbalance layer is given by 
\begin{align}
x_{c,o}[n]=\mu x_{c,i}[n]+\nu x_{c,i}^*[n],
\end{align}
where $(.)^*$ corresponds to the complex conjugate, and $(\mu,\nu)\in\mathbb{C}^2$ are the IQ imbalance parameters~\cite{TAR05}. This layer depends on the 4 real-valued parameters $\boldsymbol \theta_l=[\Re e(\mu),\Re e(\nu),\Im m(\mu), \Im m(\nu)]^T$.
\item FIR Layer: The output of a FIR layer is given by
\begin{align}
x_{c,o}[n]=\sum_{k=0}^{D-1}h[k] x_{c,i}[n-k].
\end{align}
This layer depends on the real-valued vector $\boldsymbol \theta_l=[\Re e(\mathbf{h}^T),\Im m(\mathbf{h}^T)]^T$ where $\mathbf{h}=[h[0],\cdots,h[D-1]]^T$ corresponds to the channel impulse response.
\item Piecewise-Constant Phase Layer: The output of this layer is given by 
\begin{align}
x_{c,o}[n]=e^{j\varphi[n~\text{mod} N_s]} x_{c,i}[n]
\end{align}
where {\it{mod}} corresponds to the modulo operator~\cite{DUR12}. This layer depends on $N/N_s$ real-valued parameters that are given by $\boldsymbol \theta_l=[\varphi[0],\cdots,\varphi[L-1]]^T$.
\end{itemize}

\subsection{Network training/operating}

To train the network, we propose to minimize the custom regularized loss function given by
\begin{align}
    \min_{\boldsymbol \theta,\mathbf{u},\sigma_s^2}\frac{1}{2N}\|\mathbf{y}-\mathbf{H}(\boldsymbol \theta)\mathcal{S}_{\sigma_s}(\mathbf{f}(\mathbf{x};\mathbf{u}))\|^2_2+\lambda \sigma_s^2,\label{metric}
\end{align}
where the first term corresponds to the Mean Squared Error (MSE), which gives an estimate of the \emph{observed noise variance}, and where the second term is a regularization term that depends on the \emph{data noise variance}. Note that the network training using the pilot symbols available at the receiver, $\mathbf{x}$, leads to the joint estimation of the channel parameters and symbol detection through the obtained vector $\widehat{\mathbf{u}}$. 

The regularization parameter $\lambda$ plays an important role to prevent the overfitting problem. Indeed, for small number of pilot samples $\mathbf{x}$, only minimizing the estimated observed noise variance tends to spread the additive noise into the vector $\mathbf{u}$. This noise contribution could have a negative effect on the Symbol Error Rate (SER), which is our main objective. Increasing $\lambda>0$ limits this phenomenon by encouraging a small \emph{data noise variance}.

\section{Simulation Results}\label{sec:3}

This section illustrates the advantages of our approach using a SISO communications system composed of 5 layers: a transmitter IQ imbalance, a FIR channel with $\mathbf{h} =[0.9+0.1j, 0.1+0.1j, 0.01+0.05j,0.02-0.003j, 0.004+0.012j]$, a receiver IQ imbalance, a time-varying Wiener phase noise with instantaneous phase $\varphi[n]=\sum_{k=0}^{n}\psi[n]$ where $\psi[n]\sim \mathcal{N}(0,\sigma_p^2)$, and a Gaussian additive noise $b[n]\sim \mathcal{N}(0,\frac{\sigma_b^2}{2})$. The simulation parameters are provided in Table~\ref{table0}. The transmitted signal $\mathbf{s}$ is composed of $N=200$ symbols generated from a 16QAM constellation. 

To jointly estimate $\boldsymbol \theta$ and detect $\mathbf{u}$, we consider a multi-layer parametric network composed of 6 layers: a data-augmented layer with $N_u=180$ complex-valued data symbols, a soft projector with initial parameter $\sigma_s^2=1$, a transmitter IQ imbalance layer initialized with $\boldsymbol \theta=[1,0,0,0]$, a FIR layer initialized with $\mathbf{h} =[1,0, 0, 0, 0]$, a receiver IQ imbalance layer initialized with $\boldsymbol \theta=[1,0,0,0]$, and a Piecewise-Constant Phase Layer initialized with $N/N_s=20$ zeros. Therefore, the network is composed of $1+2\times 180+4+2\times 5+4+20=399$ real-valued trainable parameters. The proposed network has been implemented using PyTorch\footnote{The source code is available at \url{https://github.com/vincentchoqueuse/ParamNET}}. The input vector $\mathbf{x}$ is composed of $N_x = 20$ complex-valued symbols transmitted periodically (one pilot symbol followed by 9 unknown data symbols). The loss function is given by \eqref{metric} and the regularization parameter is set to $\lambda=0.001$. In each experiment, the training stage uses the ADAM optimizer with a learning rate of $10^{-3}$~\cite{KIN14}. The performance of the network is evaluated by computing the SER for the complex-valued estimated symbols $\widehat{u}_c[n]=\widehat{u}[n]+j\widehat{u}[n+N_u]$ after projection on the alphabet set~$\mathcal{M}$, where $\widehat{u}[n]$ corresponds to the $n^{th}$ trainable parameter of the first layer.

\begin{table}[!t]
\centering
\begin{tabular}{ccc}
\hline
Layer & Type & Parameters \\
\hline
1 & IQ Tx & $(\mu,\nu)=(0.9-0.4j,0.4+0.1j)$\\
2 & FIR Channel & $\mathbf{h}$\\
3 & IQ Rx & $(\mu,\nu)=(1.8+0.13j,0.1+0.2j)$\\
4 & Phase Noise & $\sigma_p^2=0.000125$\\
5 & Noise & $\sigma_b^2$\\
\hline
\end{tabular}
\caption{Communication Chain Parameters.}\label{table0}
\end{table}

\begin{table}[!t]
\centering
\begin{tabular}{ccc}
\hline
Method & Activation Function & $N_{it}$ \\
\hline
Simple & $\times$ & $\times$\\
PG$\_$500 & $\times$ & 500 iterations\\
PG$\_$1000 & $\times$ & 1000 iterations\\
PG$\_$2000 & $\times$ & 2000 iterations\\
Proposed& soft thresholding & $\times$\\
\hline
\end{tabular}
\caption{Considered Techniques}\label{table1}
\end{table}

\subsection{Single Trial Analysis}
Fig.~\ref{fig3} shows the evolution of $\sigma_s^2$, the MSE and the SER during the training stage. The performance of the proposed technique is compared with the performance obtained with the same parametric network without activation layer. While the MSE is clearly lower for the simple parametric network, the SER shows that this network strongly overfits the data due to the small number of pilots. After 20000 iterations, the SER is equal to $0.0361$ for the parametric network, while the proposed network leads to a SER equal to $0.0027$.
\begin{figure}[!t]
\begin{subfigure}{.5\textwidth}
\centering
\begin{tikzpicture}
\begin{semilogyaxis}[
	width=0.95\linewidth,
	height=0.45\linewidth,
	    xmin=0,
	    ymin = 0.1,
    xmax=20000,
    xlabel={Number of iterations},
    xticklabel style={
        /pgf/number format/fixed,
        /pgf/number format/precision=4
    },
    xtick={5000,10000,15000,20000},
    scaled x ticks=false,
    grid=both,
    ylabel={$\sigma_s^2$}]
    \addplot [blue,thick] table[x index=0, y index=1,col sep=comma] {./data/one_shot/one_shot.csv};
\end{semilogyaxis}
\end{tikzpicture}
\caption{Evolution of the soft thresholding parameter $\sigma_s^2$}
\end{subfigure}
\vskip 1em
\begin{subfigure}{.5\textwidth}
\centering
\begin{tikzpicture}
\begin{semilogyaxis}[
	width=0.95\linewidth,
	height=0.45\linewidth,
    xlabel={Number of iterations},
    xmin=0,
    xmax=20000,
    grid=both,
    xticklabel style={
        /pgf/number format/fixed,
        /pgf/number format/precision=4
    },
    xtick={5000,10000,15000, 20000},
    scaled x ticks=false,
    ylabel={MSE (training)}]
    \addplot [blue,thick] table[x index=0, y index=2,col sep=comma] {./data/one_shot/one_shot.csv};
    \addplot [red,thick] table[x index=0, y index=3,col sep=comma] {./data/one_shot/one_shot.csv};
     \legend{No activation, soft thresholding}
\end{semilogyaxis}
\end{tikzpicture}
\caption{Evolution of the MSE (training performance)}
\end{subfigure}
\vskip 1em
\begin{subfigure}{.5\textwidth}
\centering
\begin{tikzpicture}
\begin{semilogyaxis}[
	width=0.95\linewidth,
	height=0.45\linewidth,
	xmin=0,
	ymin=0.001,
	ymax=1,
    xmax=20000,
    xlabel={Number of iterations},
    grid=both,
    xticklabel style={
        /pgf/number format/fixed,
        /pgf/number format/precision=4
    },
    xtick={5000,10000,15000, 20000},
    scaled x ticks=false,
    %legend pos=south west,
    ylabel={SER (testing)}]
    \addplot [blue,thick] table[x index=0, y index=4,col sep=comma] {./data/one_shot/one_shot.csv};
    \addplot [red,thick] table[x index=0, y index=5,col sep=comma] {./data/one_shot/one_shot.csv};
    \legend{No activation, soft thresholding}
\end{semilogyaxis}
\end{tikzpicture}
\caption{Evolution of the Symbol Error Rate (SER) (testing performance)}
\end{subfigure}
\caption{Evolution of $\sigma_s^2$ and SER performance (SNR=$20$dB).}\label{fig3}
\end{figure}
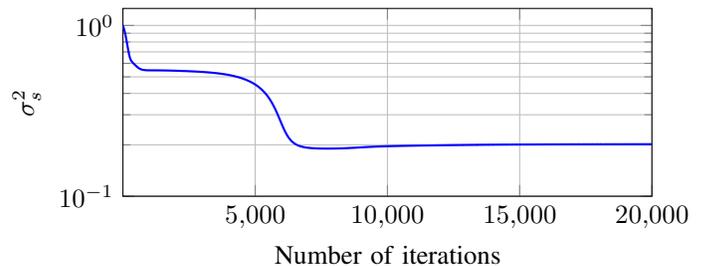
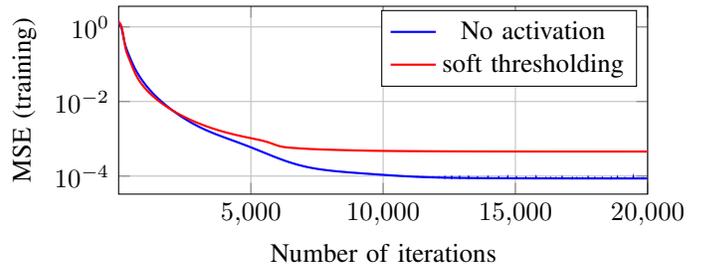
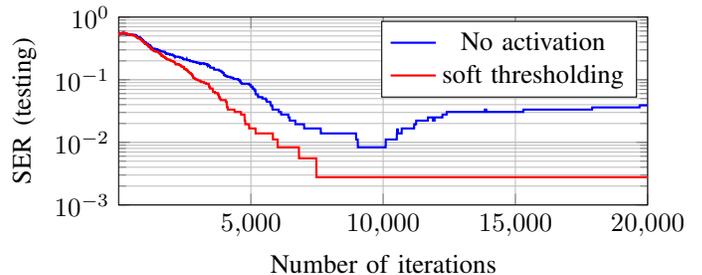
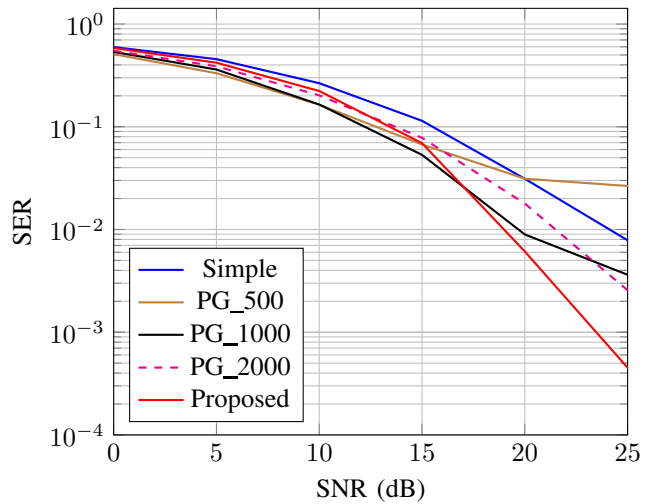
\begin{figure}[!t]
\centering
\begin{tikzpicture}
\begin{semilogyaxis}[
	width=0.95\linewidth,
    xlabel={SNR (dB)},
    grid=both,
    legend pos=south west,
    xmin=0,
    xmax=25,
    ymin=0.0001,
    ylabel={SER}]
    \addplot [blue,thick] table[x index=0, y index=1,col sep=comma] {./data/monte_carlo/ser.csv};
    \addplot [thick,brown] table[x index=0, y index=2,col sep=comma] {./data/monte_carlo/ser.csv};
    \addplot [thick,black] table[x index=0, y index=3,col sep=comma] {./data/monte_carlo/ser.csv};
    \addplot [thick,magenta,dashed] table[x index=0, y index=4,col sep=comma] {./data/monte_carlo/ser.csv};
    \addplot [thick,red] table[x index=0, y index=5,col sep=comma] {./data/monte_carlo/ser.csv};
    \legend{Simple, PG$\_$500 ,PG$\_$1000,PG$\_$2000,Proposed  }
\end{semilogyaxis}
\end{tikzpicture}
\caption{Testing performance versus SNR.}\label{fig4}
\end{figure}

\subsection{Monte Carlo Simulations}

Fig.~\ref{fig4} presents the average SER versus SNR using Monte Carlo simulations. In each Monte Carlo trial, we generate a new realization of the Wiener phase and additive noises, and the optimization is stopped after $N_{it}=10000$ iterations. The performance of the proposed technique is compared with the performances of the techniques reported in Table \ref{table1}. 
The Projected Gradient (PG) techniques are obtained by periodically projecting the trainable parameters $\mathbf{u}$ into the set $\mathcal{M}$ after $N_{it}$ iterations~\cite{TAK19}. As the projection rate has a significant impact on the SER performance, we have considered several PG detectors with different values of $N_{it}$. We observe that the simple technique requires a large SNR to perform well. Furthermore, we also note that the performance of the PG techniques critically depends on the projection rate and SNR. In particular, projecting frequently $\mathbf{u}$ into the set $\mathcal{M}$ leads to poor performance for medium or high SNR. Regarding the proposed method, it leads to the best performance at medium and high SNR.

\section{Conclusion}

This paper proposes a new network architecture called ParamNet for joint estimation and data detection in communications systems. ParamNet is composed of a data augmented layer and a soft thresholding layer that encourages the data constellation. This network can be trained using a small number of pilot symbols and a custom regularized loss function. Simulation results have shown that this strategy outperforms projected gradient techniques. In particular, the proposed strategy provides a solution to avoid the overfitting problem in the medium and high SNR regions. Future works will investigate the application of the proposed network in more complex communications chains, including MIMO systems.
\bibliographystyle{IEEEtran}  
\bibliography{biblio/biblio.bib}

\end{document}